\documentclass{article}

\usepackage{microtype}
\usepackage{graphicx}
\usepackage{subfigure}
\usepackage{booktabs} 

\usepackage{hyperref}
\usepackage{xspace}

\newcommand{\pt}{\ensuremath{p_\mathrm{T}}\xspace}

\usepackage[accepted]{latinx-icml-2021}

\icmltitlerunning{LXAI 2021}

\begin{document}

\twocolumn[
\icmltitle{Sparse Data Generation for Particle-Based\\
        Simulation of Hadronic Jets in the LHC}



\icmlsetsymbol{equal}{*}

\begin{icmlauthorlist}
\icmlauthor{Breno Orzari}{spr}
\icmlauthor{Thiago R. F. P. Tomei}{spr}
\icmlauthor{Maurizio Pierini}{cern}
\icmlauthor{Mary Touranakou}{cern,nkua}
\icmlauthor{Javier Duarte}{ucsd}
\icmlauthor{Raghav Kansal}{ucsd}
\icmlauthor{Jean-Roch Vlimant}{caltech}
\icmlauthor{Dimitrios Gunopulos}{nkua}
\end{icmlauthorlist}

\icmlaffiliation{cern}{European Organization for Nuclear Research (CERN)}
\icmlaffiliation{spr}{SPRACE-Unesp, S\~{a}o Paulo, Brazil}
\icmlaffiliation{ucsd}{University of California San Diego, La Jolla, California, United States of America}
\icmlaffiliation{caltech}{California Institute of Technology, Pasadena, California, United States of America}
\icmlaffiliation{nkua}{National and Kapodistrian University of Athens, Athens, Greece}

\icmlcorrespondingauthor{Breno Orzari}{breno.orzari@cern.ch}

\icmlkeywords{Machine Learning, ICML, High Energy Physics}

\vskip 0.3in
]



\printAffiliationsAndNotice{} 

\begin{abstract}
We develop a generative neural network for the generation of sparse data in particle physics using a permutation-invariant and physics-informed loss function. The input dataset used in this study consists of the particle constituents of hadronic jets due to its sparsity and the possibility of evaluating the network's ability to accurately describe the particles and jets properties. A variational autoencoder composed of convolutional layers in the encoder and decoder is used as the generator. The loss function consists of a reconstruction error term and the Kullback-Leibler divergence between the output of the encoder and the latent vector variables. The permutation-invariant loss on the particles' properties is combined with two mean-squared error terms that measure the difference between input and output jets mass and transverse momentum, which improves the network's generation capability as it imposes physics constraints, allowing the model to learn the kinematics of the jets. 
\end{abstract}

\section{Introduction}
\label{submission}
The Large Hadron Collider (LHC) at CERN~\cite{evans:2008zzb}
is a proton-proton (pp) collider that allows fundamental physics research at the highest energy regimes.
Fast simulation of high energy physics (HEP) objects for particle physics data analysis has been a challenge for the last decades; the problem became exacerbated in the LHC environment. 
With the advent of machine learning (ML) techniques being applied to event reconstruction, jets classification, and other necessities of experimental particle physics~\cite{Guest_2018,hepml}, the usage of ML to build a generator of events can be seen as the next step for the current Monte Carlo based generators. 
Although it is a challenging task, it can be split into the generation of the distinct final-state objects of particle collisions.
For the high energy pp collisions that take place at the LHC, the most common kinds of particles produced are \emph{hadrons} that consist of quarks and gluons, such as the proton and neutron.
Hadrons produced in these collisions usually appear in collimated groups, called \emph{hadronic jets}~\cite{subjettiness2,jet_reco}.

We measure the jet physical properties by setting sensible detector elements around the pp collision region.
Those elements allow the measurement of the energy and momenta (for charged particles) of the jet constituents.
Those combined measurements are used to determine the jet characteristics through the particle-flow (PF) algorithm ~\cite{PFcms,PFatlas}. 
The jet can then be described as a sparse, unordered set of constituent particles where each particle is further characterized by its properties like energy, momentum\footnote[2]{The coordinate system used in HEP experiments is as follows:
origin set at the center of the local pp collision region;
$z$ axis along the beam direction,
$y$ axis vertically upward.
3D coordinates are usually given in terms of
$\rho = (x^2 + y^2)^{1/2}$, azimuth $\phi$ and
pseudorapidity $\eta = -\ln \tan (\theta/2)$, where $\theta$ is the polar angle.}, charge, particle type, among others. 
To deal with the sparsity of the data, one might use graph neural networks~\cite{Duarte:2020ngm,Shlomi:2020gdn} without assuming any specific ordering. 
For the same reason, energy flow networks~\cite{Komiske_2019}, physics-inspired permutation-invariant architectures, were introduced. 

In general, the loss function used to train such models is the mean squared error (MSE). However, using such a loss term implies breaking the permutation invariance of the data. To avoid such behavior, in this paper we use a loss function based on a nearest-neighbor distance (NND) or Chamfer distance~\cite{10.5555/1622943.1622971,Fan_2017_CVPR}.
The aim of this work is to develop a generative neural network capable of producing sets of non-ordered particles as constituents of hadronic jets which correctly reproduce the particles and jets' physical properties with high fidelity, using a permutation-invariant reconstruction error term in the loss function that, for this purpose, has not been applied so far.

This paper is organized as follows: we introduce the benchmark dataset and model in Section~\ref{sec:data}. 
The loss function and evaluation metric used are described in Section~\ref{sec:loss}. 
We show performance on applications in Section~\ref{sec:application}.  
Conclusions and future steps are given in Section~\ref{sec:conclusions}.

\section{Benchmark Dataset and Model}
\label{sec:data}

The input dataset\footnote[3]{\href{https://zenodo.org/record/3602254}{100 particles dataset}} 
consists of lists of 100 particles that are constituents of hadronic jets. 
Where less than 100 particles are present in the jet, the list is filled with zero-padded entries for the remaining particles up to 100. 
Each particle is described by its momentum components as $\vec{p} \ = \ (p_{x}, p_{y}, p_{z})$ in units of giga-electron volts (GeV). 
Approximately 177,000 examples of high-momentum jets originating from gluons\footnote[4]{The dataset also contains simulated jets that originate from W bosons, Z bosons, top quarks (t), and light quarks (q), that were not used for this work, but will be used for future research.} were generated, and this data is split 70\% for training, 15\% for validation and 15\% for testing. 
A feature-dependent normalization is applied in the data such that the range of each particle feature is [0.0, 1.0].

\begin{figure*}[h]
    \centering
    \includegraphics[width=0.8\textwidth]{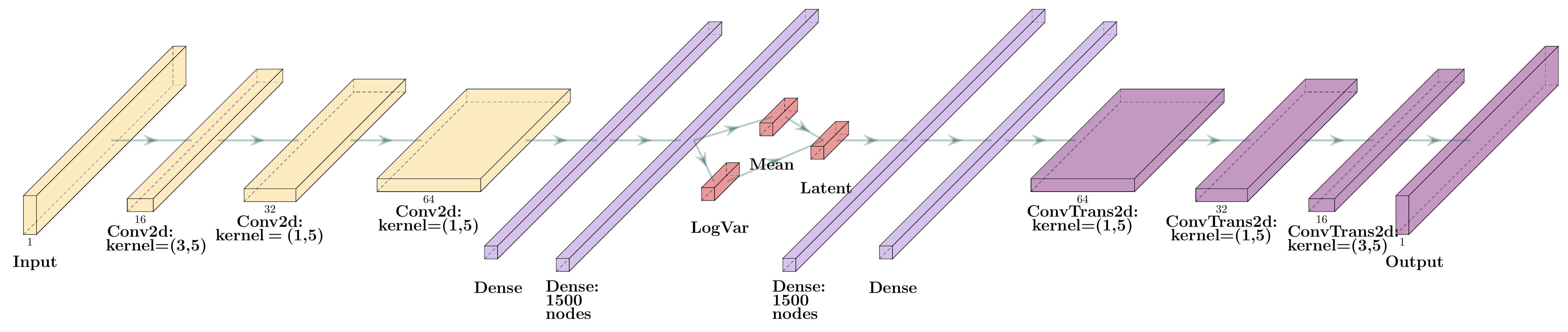}
   \caption{Schematic representation of the VAE architecture used in this study. 
   For each convolutional (dense) layer the kernel size (number of neurons) is specified. \label{fig:arch}}
\end{figure*}

A variational autoencoder (VAE) is used as the generative neural network model, composed of 3 convolutional layers and 2 dense layers in the encoder, a latent vector of dimension 20 as well as 2 dense layers and 3 convolutional layers in the decoder. 
After each layer the rectified linear unit (ReLU) 
 activation function is applied, except after the last encoder layer, that is used for the reparameterization, and the last decoder layer, in which the sigmoid 
 activation function is used to constrain the output of the network to be between [0.0, 1.0]. 
 A schematic representation of the architecture used is depicted in figure \ref{fig:arch}.

\section{Error Function and Evaluation Metric}
\label{sec:loss}

In the training step, the main purpose of the VAE is to reconstruct the input data as closely as possible, while encoding the information in the latent vector for the generation of new data. 
The product of this step is referred to as the output dataset, which is composed of sets of reconstructed particles described by $\vec{\hat{p}} \ = \ (\hat{p}_{x}, \hat{p}_{y}, \hat{p}_{z})$. 
To achieve this goal, the loss function can be written as a reconstruction error between input and output data, and a Kullback-Leibler (KL) divergence ($D_\mathrm{KL}$) term, which forces the probability distribution of the latent vector values to take the form of a simple distribution, such as a standard Gaussian one. 
The loss function is then written as
\begin{equation}
    L_\mathrm{VAE} = (1-\beta)L_\mathrm{rec}  + \beta D_\mathrm{KL},
    \label{eq:lossvae}
\end{equation}
where the $\beta$ parameter~\cite{Higgins2017betaVAELB} is a relative weighting factor between $L_\mathrm{rec}$ and $D_\mathrm{KL}$.

The reconstruction term is further composed by 3 distinct contributions. 
The first is an NND term that quantifies the difference between input and output particles properties in a permutation-invariant way, following the expression
\begin{equation}
D^\mathrm{NND}(\mathcal{J}_{k}, \hat{\mathcal{J}}_{k}) =  \sum_{i \in \mathcal{J}_{k}} \min_{j \in \hat{\mathcal{J}}_{k}} D(\vec{p}_i, \vec{\hat{p}}_j) + \sum_{j \in \hat{\mathcal{J}}_{k}} \min_{i \in \mathcal{J}_{k}} D(\vec{p}_i, \vec{\hat{p}}_j),
\label{eq:D_NND}
\end{equation}
where $D$ is the squared Euclidean distance, $\mathcal{J}$ and $\hat{\mathcal{J}}$ represent the \textit{kth} input and output jets, respectively, and the indices $i$ and $j$ represent the \textit{ith} and \textit{jth} particles inside a given jet. Summing over all the jets in the dataset, we have
\begin{equation}
L^\mathrm{NND} =  \sum_{k} D^\mathrm{NND}(\mathcal{J}_{k}, \hat{\mathcal{J}}_{k}).
\label{eq:L_NND}
\end{equation}

The other two terms measure the difference between input and output jets' \pt and invariant mass, thus forcing the network to learn the jets characteristics as well.
They are written as
\begin{equation}
L^\mathrm{J} = \sum_{k} \left[ \gamma_{\pt}\mathrm{MSE}(p_{\mathrm{T}k}, \hat{p}_{\mathrm{T}k}) + \gamma_{m}\mathrm{MSE}(m_{k}, \hat{m}_{k})~ \right],
\label{eq:L_J}
\end{equation}
where $p_{\mathrm{T}k}$ ($\hat{p}_{\mathrm{T}k}$) and $m_{k}$ ($\hat{m}_{k}$) are the \textit{kth} input (output) jet \pt and mass, respectively, and $\gamma_{\pt}$ and $\gamma_{m}$ are weights that can be optimized. Here it would be important to highlight that, since the dataset is composed of the particles features only, the calculation of the jets characteristics from the particles properties needs to be performed for every input/output jet. However, to properly compute these quantities, the normalization applied in the dataset has to be inverted, because the sum of the particles' momenta in absolute GeV scale are necessary.

At the generation step, a set of random values sampled from a 
Gaussian distribution is fed into the VAE latent vector, and the network output will be referred to as the generated dataset. The earth mover's (Wasserstein) distance (EMD)~\cite{emd} has been also proposed as a permutation-invariant metric to compare unordered points in sets~\cite{Fan_2017_CVPR}, however its use directly in the loss function has been shown to be computationally costly. Hence, we use EMD to measure the generating capabilities of the network by calculating the distance between the 1D probability distributions of input and generated jets mass, \pt, energy, $\eta$ and $\phi$. The sum of these 5 quantities, referred to as EMD$_\mathrm{sum}$, is used to measure the network's performance.

\section{Applications}
\label{sec:application}

The architecture is implemented in PyTorch
, and trained using the early stopping technique~\cite{JMLR:v15:raskutti14a,matet2017dont} at first, with a patience of 15 epochs, and at a second stage it was trained for 1,500 epochs (approximately 4--6 times the number of epochs as before) to test the effect of a longer training period. 
The optimizer used is Adam~\cite{adam} with a learning rate of 0.0001. 
The parameter $\beta$ was set to 0.9998, due to the large values of $L_\mathrm{rec}$ compared to $D_\mathrm{KL}$, but there is room for further optimization. 
The hyperparameter $\gamma_{\pt}$ was set to 1.0 and remained unchanged, while $\gamma_{m}$ was set as 1.0 at first, and was increased to 10.0 later, since, it showed improvements in the generative capabilities of the network.

Every 50 epochs, each of the trained models was set to generate around 26,000 jets for the evaluation of the generating properties of the network. 
The EMD metric was calculated and the best model for each training section was selected as the one that showed the smallest EMD$_\mathrm{sum}$. Table~\ref{table:emdsum} shows the results of four distinct models trained as described above: using early stopping with $\gamma_{m}$ = 1.0 (ES1) and $\gamma_{m}$ = 10.0 (ES10); training for more epochs with $\gamma_{m}$ = 1.0 (ME1) and $\gamma_{m}$ = 10.0 (ME10).

\begin{table}[h]
\caption{Best EMD$_\mathrm{sum}$ values for distinct models: early stopping with $\gamma_{m}$ = 1.0 (ES1) and $\gamma_{m}$ = 10.0 (ES10); training for more epochs with $\gamma_{m}$ = 1.0 (ME1) and $\gamma_{m}$ = 10.0 (ME10)}
\label{table:emdsum}
\vskip 0.15in
\begin{center}
\begin{small}
\begin{sc}
\begin{tabular}{lr}
\toprule
Model & EMD$_\mathrm{sum}$ \\
\midrule
ES1 & 0.0119\\
ME1 & 0.0090\\
ES10 & 0.0085\\
\textbf{ME10} & \textbf{0.0062}\\
\bottomrule
\end{tabular}
\end{sc}
\end{small}
\end{center}
\vskip -0.1in
\end{table}

Based on the EMD$_\mathrm{sum}$, we observe a large improvement in the VAE generation of hadronic jets when increasing both the $\gamma_{m}$ parameter and the number of epochs. 
The reason for the former is that the largest contribution to the evaluation metric comes from the distinction between input and generated jets mass, and, making the MSE on the jets mass term more important in the error, resulted in a better generation of the jets mass.

\begin{figure}[h]
    \centering
    \includegraphics[width=0.2\textwidth]{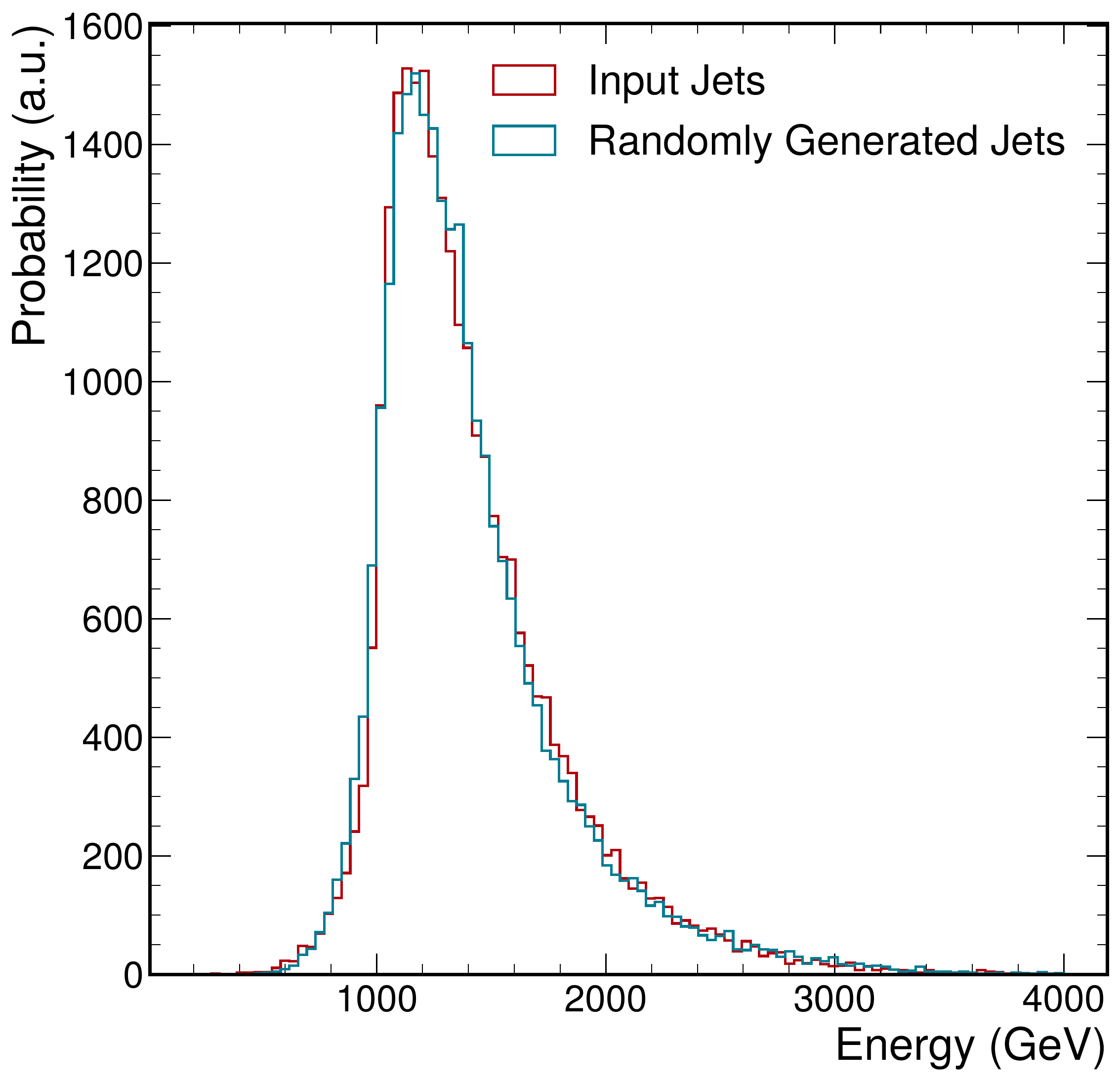}
    \includegraphics[width=0.2\textwidth]{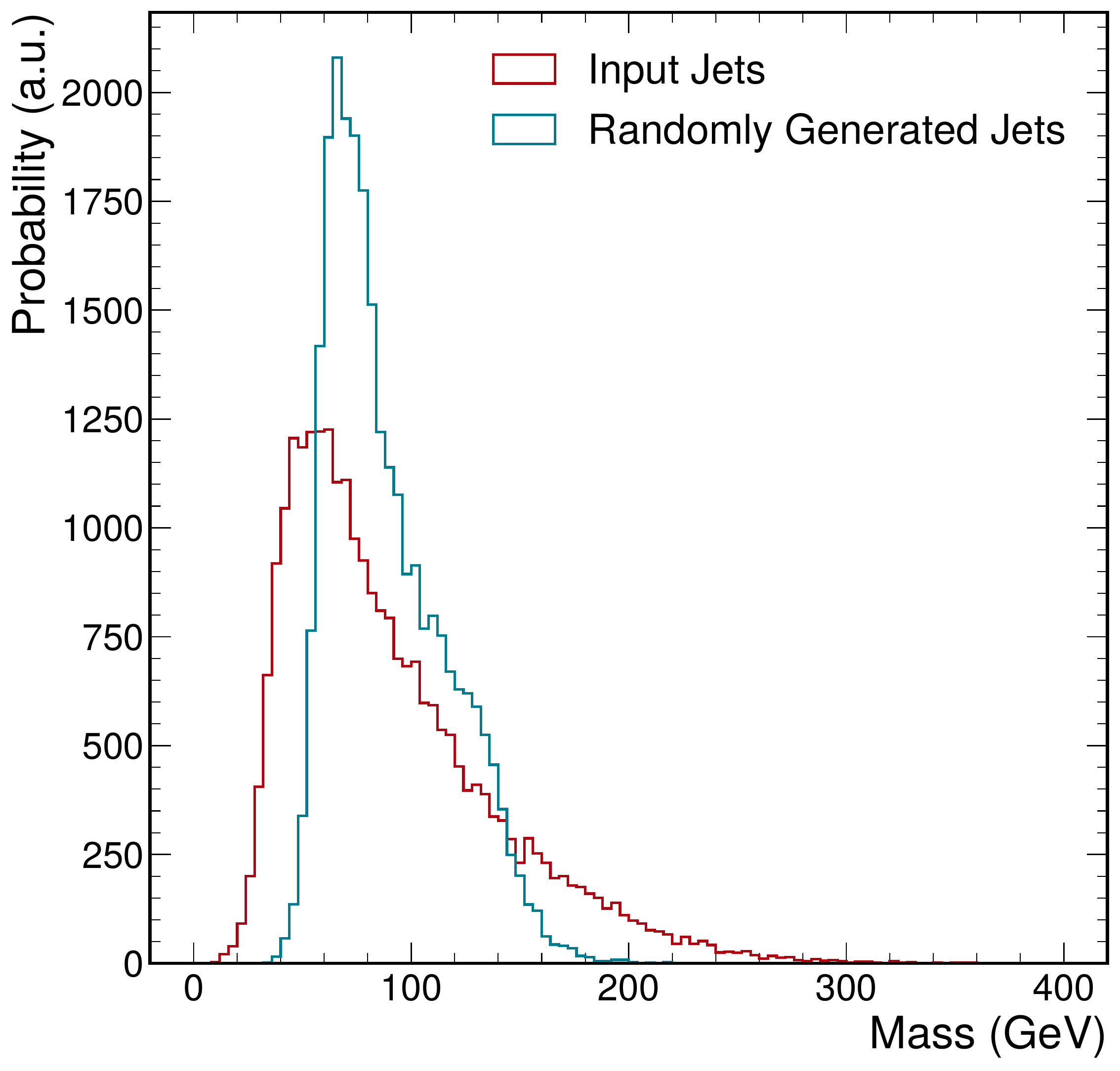}
    \includegraphics[width=0.2\textwidth]{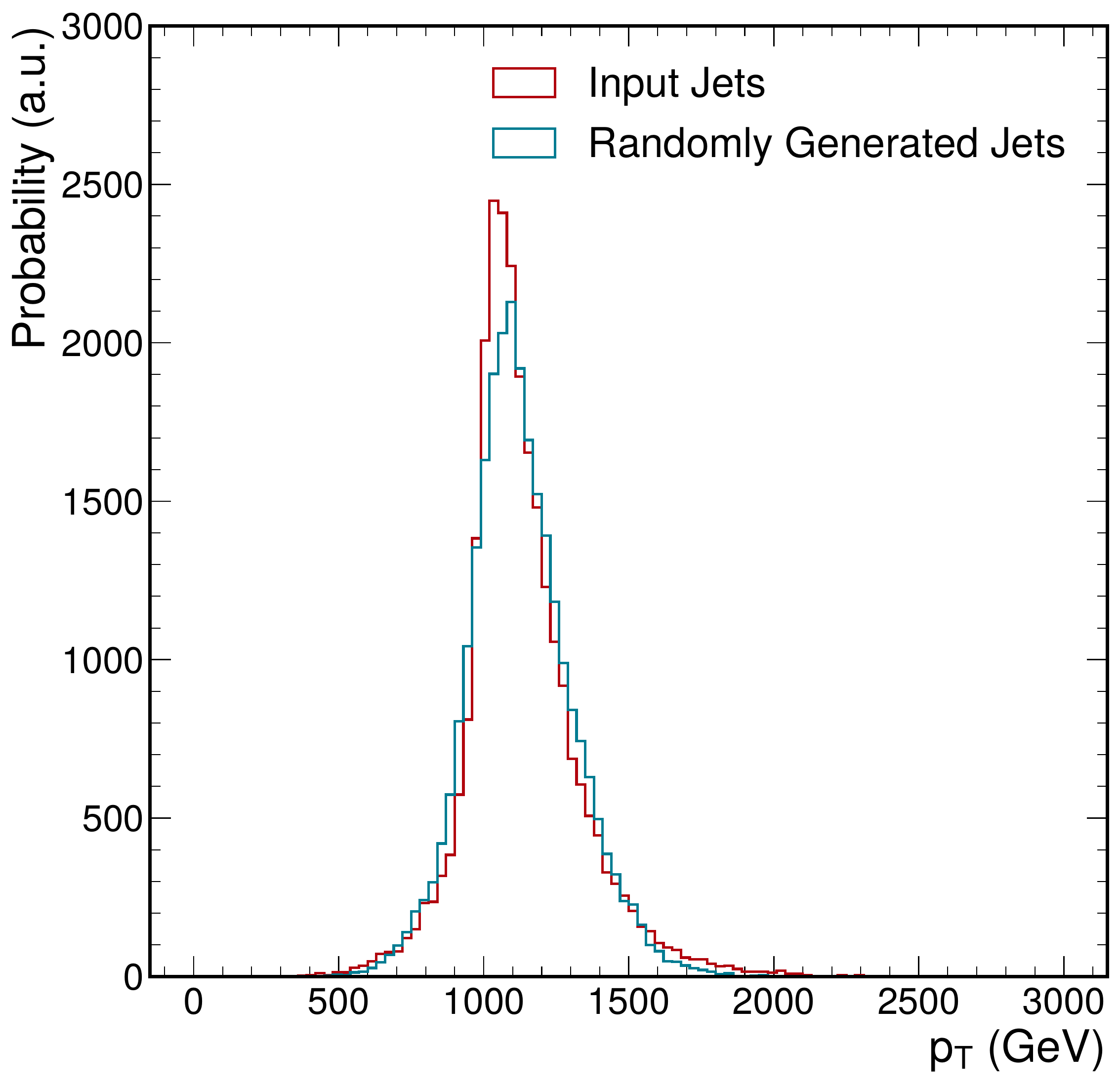}
    \includegraphics[width=0.2\textwidth]{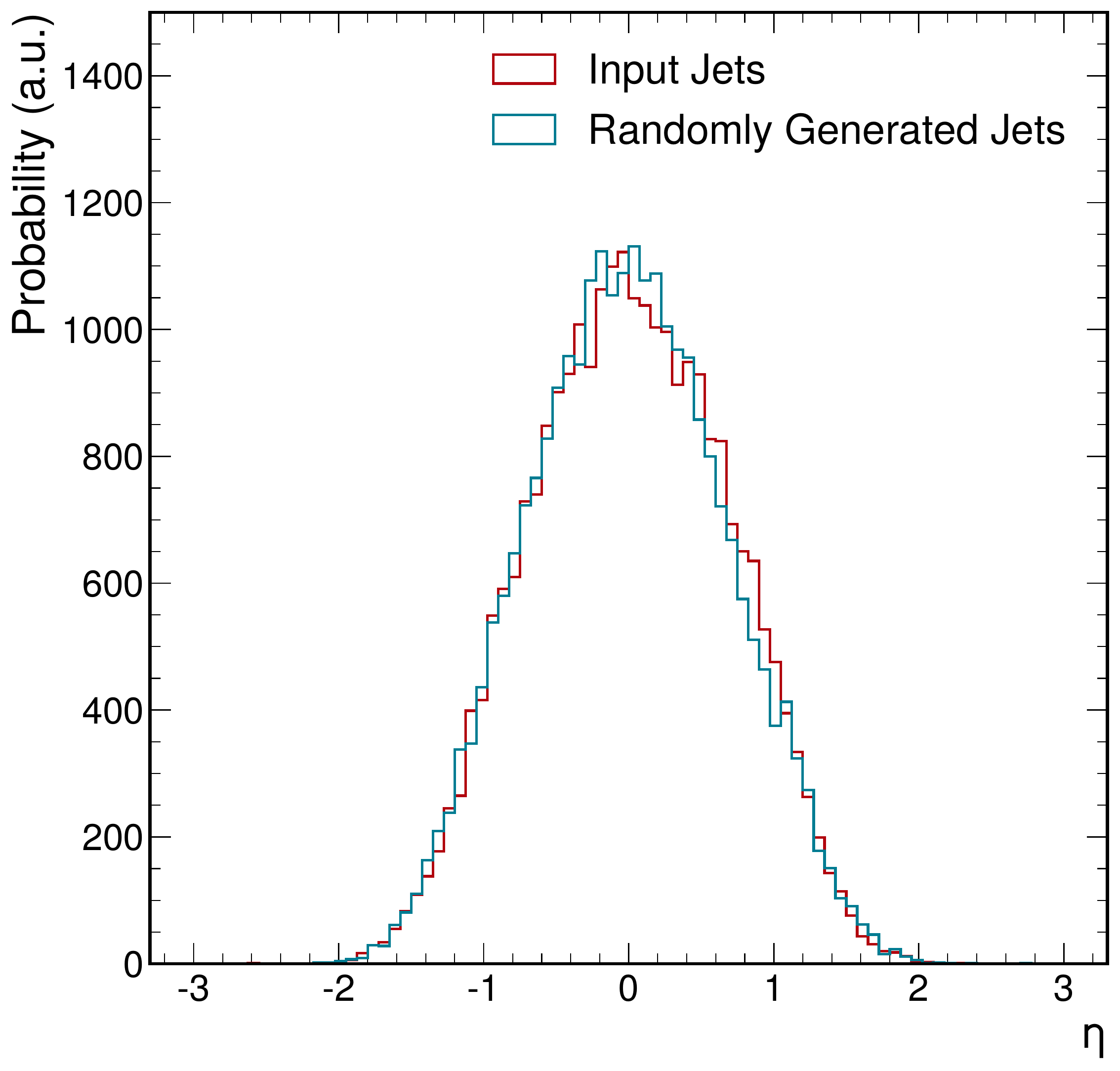}
   \caption{Input (red) and generated (blue) jets variables for the best performing model (ME10). 
   Top: energy (left), mass (right). Bottom: \pt (left), $\eta$ (right). \label{fig:jetvar}}
\end{figure}

Figure~\ref{fig:jetvar} shows the distributions of some input and generated jets variables for the best model performance (ME10). 
Although the input mass distribution is not reproduced by the generated jets, all of the other generated jet variables have similar input and generated distributions.

\section{Summary and Outlook}
\label{sec:conclusions}

We presented a generative neural network suited for the generation of sparse data, showing an application of a VAE for the generation of hadronic jets, just like the ones that are collected as data at the LHC. 
Distinct techniques for the training step of the VAE were applied, and the best model showed a value of the EMD$_\mathrm{sum}$ metric of 0.0062. 
Although the comparison of input and generated jets mass histograms shows a difference in its distributions, other relevant variables showed good performance. 
There is still a lot of room for improvement through the optimization of the network hyperparameters. 
In order to improve the agreement and obtain a more accurate jet generator, we plan to study the addition of normalizing flows~\cite{normalizing_flows} in the latent space to learn a better posterior that what we obtain from Gaussian sampling.

\section*{Acknowledgements}
B.~O. and T.~T. are supported by grant \#2018/25225-9, SãoPaulo Research Foundation (FAPESP). B.~O. was also partially supported by grants \#2018/01398-1 and \#2019/16401-0, São Paulo Research Foundation (FAPESP).
M.~T. and M.~P. are supported by the European Research Council (ERC) under the European Union’s Horizon 2020 research and innovation program (Grant Agreement No. 772369).
J-R.~V. is supported by the U.S. Department of Energy (DOE), Office of Science, Office of High Energy Physics under Award No. DE-SC0011925, DE-SC0019227, and DE-AC02-07CH11359.
J.~D. and R.~K. are supported by the DOE, Office of Science, Office of High Energy Physics Early Career Research program under Award No. DE-SC0021187 and by the DOE, Office of Advanced Scientific Computing Research under Award No. DE-SC0021396 (FAIR4HEP).

\bibliography{main}
\bibliographystyle{icml2021}

\end{document}